# scMTNI: Leveraging cellular trajectory and context to infer dynamic GRNs from single-cell multi-omics data

**Suvojit Hazra[1], Chandrani Kumari[1], Sushmita Roy[1,2]**

[1]Wisconsin Institute for Discovery, University of Wisconsin-Madison, Madison, WI, USA
[2]Department of Biostatistics and Medical Informatics, University of Wisconsin-Madison, Madison, WI, USA

## Abstract

Transcriptional gene regulatory networks (GRNs) depict the directed relationships between regulators and target genes, determining gene expression patterns in a cell-type-specific manner. Single-cell multi-omics technologies, such as single-cell RNA sequencing (scRNA-seq) and single-cell Assay for Transposase-Accessible Chromatin using sequencing (scATAC-seq), enable high-resolution measurement of cell-type-specific gene expression and regulation in an unprecedented way. However, tools for inferring cell-type-specific GRNs and modeling their dynamics remain scarce. To facilitate the inference and analysis of cell-type-specific GRNs in contexts such as cellular development or disease progression, where cell lineage structure and dynamics are important, we developed a multi-task learning framework, single-cell Multi-Task Network Inference (scMTNI). scMTNI and its associated network analyses tools offer a comprehensive package to define cell-type-specific GRNs and examine their dynamics. This book chapter describes the scMTNI tool and demonstrates its application to an existing cellular reprogramming single cell multi-modal dataset to infer cell-type-specific GRNs and identify key regulators of cellular fate transitions during cellular reprogramming.

# 1. Introduction

Gene regulatory networks (GRNs) define the connections between regulatory proteins such as transcription factors and target genes, specifying the spatiotemporal expression patterns of genes in diverse cellular contexts, thereby shaping cellular identity and function [1]. GRNs reconfigure to modulate cell-type-specific gene expression levels during cellular development [2, 3], tissue regeneration [4], disease progression [5], plant developmental and stress response processes [6], and evolutionary adaptation [7], and their rewiring underlies many cellular transitions. An important challenge in biology is to map GRNs at high resolution and understand how they change across different cell states. Recent multi-omics measurements, such as single-cell RNA sequencing (scRNA-seq) and single-cell Assay for Transposase-Accessible Chromatin using sequencing (scATAC-seq), offer high resolution for discovering cell-type-specific GRN dynamics [8]. However, computational tools that leverage these measurements to define the GRNs driving cell-type-specific gene expression patterns are still in their early stages.

To address this gap, we developed single-cell Multi-Task Network Inference (scMTNI) [9], a multi-task learning framework that integrates multi-omics data, scRNA-seq and scATAC-seq, with cell lineage structure to jointly infer cell-type-specific gene regulatory networks (GRNs). scMTNI models GRNs as a probabilistic graphical model and uses multi-task learning to capture shared and distinct regulatory programs across related cell types, treating each cell type or cluster as a distinct but related task. The cell lineage hierarchy provides a probabilistic prior that constrains and guides the network inference. scMTNI can be applied to datasets with or without matched scATAC-seq data, but in both cases, it constrains the inferred networks based on the sequence-specific motif information. As part of downstream analysis, scMTNI provides tools based on *k*-means edge clustering and Latent Dirichlet Allocation (LDA) topic modeling, to compare inferred GRNs across cell types to examine dynamics and rewiring of the cell-type-specific GRNs. This can be beneficial to prioritize regulators based on the extent of rewiring along the lineage.

The scMTNI package, available at https://github.com/Roy-lab/scMTNI, implements the core algorithm for inferring cell-type-specific GRNs based on cell clusters within a developmental lineage. It also includes tools for network dynamics analysis, such as *k*-means clustering and LDA topic modeling. In this chapter, we first provide the theory behind the scMTNI framework and then demonstrate its application on an existing mouse cellular reprogramming multi-omics dataset [9, 10]. This step-by-step guide is intended to help users apply scMTNI to other systems where modelling GRN dynamics across related cell states is of interest.

## 2. The theory of scMTNI

The multi-task learning framework of scMTNI uses a probabilistic model [11], where each task learns the gene regulatory networks (GRNs) for each cell cluster from the cell lineage. Each GRN is modeled as a dependency network [12], which is a class of probabilistic graphical models that represent predictive statistical relationships among random variables. Each gene is modeled as a random variable $X_j^d$, which encodes the expression level of gene $j$ in a cell type $d$. A conditional probability distribution $P(X_j^d \mid X_i^d)$ models the strength of the interaction between a regulator $i$ and a gene $j$ in the cell type $d$. In scMTNI, these distributions are estimated using conditional Gaussian distributions. scMTNI incorporates two priors: (1) a cell-type-specific prior, (2) a cell lineage prior. Below we describe these priors.

The cell-type-specific prior controls the sparsity of the network and favors a regulatory edge if a motif instance of the regulator exists in the promoter region. This motif information can be cell-type-specific if relevant scATAC-seq data are available or can be cell type agnostic. As we have done previously [13, 14], we model the prior probability of edge presence using a logistic function:

$$P\left(I_{u,v}^{(d)} = 1\right) = \frac{1}{1 + e^{-(\beta_0 + \beta_1 * m_{ij})}}$$

Here $I_{u,v}^{(d)}$=1 indicates a regulatory interaction between transcription factor $u$ and gene $v$, and $I_{u,v}^{(d)}$=0 indicates no interaction. The parameter $\beta_0$ serves as a sparsity prior, controlling the penalty associated with introducing new edges into the network. This parameter typically takes a negative value ($\beta_0 < 0$), and a smaller (i.e., more negative) value imposes a stronger penalty, encouraging the model to infer a sparser regulatory network. In addition to sparsity, prior biological knowledge from sequence-specific motifs is incorporated through the parameter $\beta_1$, which modulates the influence of motif information in the edge inference. $\beta_1$ is non-negative ($\beta_1 \geq 0$), with higher values imposing a stronger emphasis on the presence of a motif when determining whether to include an edge. When motif data is not available, $\beta_1$ is set to zero, effectively disabling motif influence in the model. The motif information itself is captured by $m_{ij}$, which denotes the motif instance score, representing the presence or strength of a binding motif for regulator $i$ within the promoter region of gene $j$. The final edge-specific prior probability is a tradeoff between the sparsity and motif prior parameters.

The second prior in scMTNI is the cell lineage tree prior, which incorporates known relationships among cell types to guide the inference of the GRNs. Rather than inferring GRNs for each cell type independently, scMTNI assumes that cell types closer on the lineage share more regulatory features, than cell types further away. The prior is encoded using three hyper parameters: (i) $p_r$: the probability that a regulatory edge exists in the root (starting) cell type, (ii) $p_g$: the probability of gaining a new regulatory edge in a child cell type if it was absent in the parent cell type; (iii) $p_m$: the probability of maintaining an existing edge in the child cell type if it was already present in the parent cell type. These three probabilities define how GRNs evolve over the lineage tree and serve as prior belief over network structure. A low value of $p_r$ results in a sparser regulatory network in the root cell type; for example, setting $p_r$=0.2 means there is only a 20% chance that any given regulatory edge is initially present in the root. A low value of $p_g$ means that gaining new regulatory edges in the descendent cell types is relatively unlikely unless strongly supported by expression and/or accessibility data; for example, setting $p_g$=0.2 would indicate a higher chance (20%) of new regulatory edges forming in descendent cells. A high value of $p_m$ (such as 0.8) means that existing regulatory connections are very likely (80% chance) to be kept as cells develop, reflecting the fact that many gene regulations stay the same between related cell types. By incorporating this prior, scMTNI captures both shared and cell type unique regulatory patterns across a lineage.

## 3. Materials

scMTNI source code developed in C++, compiled and tested for Linux environment. scMTNI is freely available from the Roy Lab GitHub repository: https://github.com/Roy-lab/scMTNI/tree/master. scMTNI compilation requires the GCC compiler (≥v6.3.1) and the GNU Scientific Library (GSL) (≥v2.6). scMTNI takes as input: (i) a normalized scRNA-seq gene expression matrix, (ii) a cell lineage tree or cluster hierarchy, and (iii) optionally, prior regulatory information derived from scATAC-seq via transcription factor (TF) motif enrichment. If scATAC-seq data is not available, scMTNI can still operate using only scRNA-seq and lineage information.

To run scMTNI on a dataset with seven cell types, approximately 12,250 genes, and 2,000 regulators, a standard laptop or desktop with at least 1 GB of available memory is sufficient, with an expected runtime of 1 hour under typical settings [9].

For each cell type, scMTNI requires the cell-type-specific expression data, regulators information, cell lineage tree, cell lineage order, a mapping file for genes across all cell types, and an optional collection of cell-type-specific regulator-gene prior network files from sequence-specific motif information mapped by accessibility profile. The following section demonstrates the steps for applying scMTNI with Prior (scMTNI+Prior), using gene expression data from six cellular stages of mouse reprogramming: MEF, FBS-Day3, FBS-Day6, FBS-Day9, FBS-Day12, and mESC (**Figure 1**). The expression data can be obtained from the publicly available dataset GSE108222 [9].

To incorporate transcription factor (TF) binding priors, sequence-specific motif and TF information can be retrieved from the CIS-BP database [15] and combined with cell-type-specific chromatin accessibility data to generate a TF-target gene prior network tailored to each cell cluster. Chromatin accessibility profiles for the same cell stages were derived from an ATAC-seq dataset (GEO ID: GSE208620 [9]), enabling motif-to-accessibility mapping for accurate assignment of regulatory edges. This integrated prior was used to guide GRN inference in scMTNI, enhancing the biological relevance of the inferred regulatory programs across mouse cellular reprogramming stages. The filtered expression dataset consists of 3,460 single cells spanning the six cellular stages, which are further distributed across eight distinct cell clusters (C1-C8). Each cluster includes expression measurements for 14,953 genes, among which 576 are annotated as regulatory transcription factors (TFs). The original versions of the expression data, motif instances, and prior networks used in the scMTNI manuscript are publicly available at Zenodo (https://zenodo.org/records/7879228). The versions used for the book chapter are available at https://zenodo.org/records/16914547.

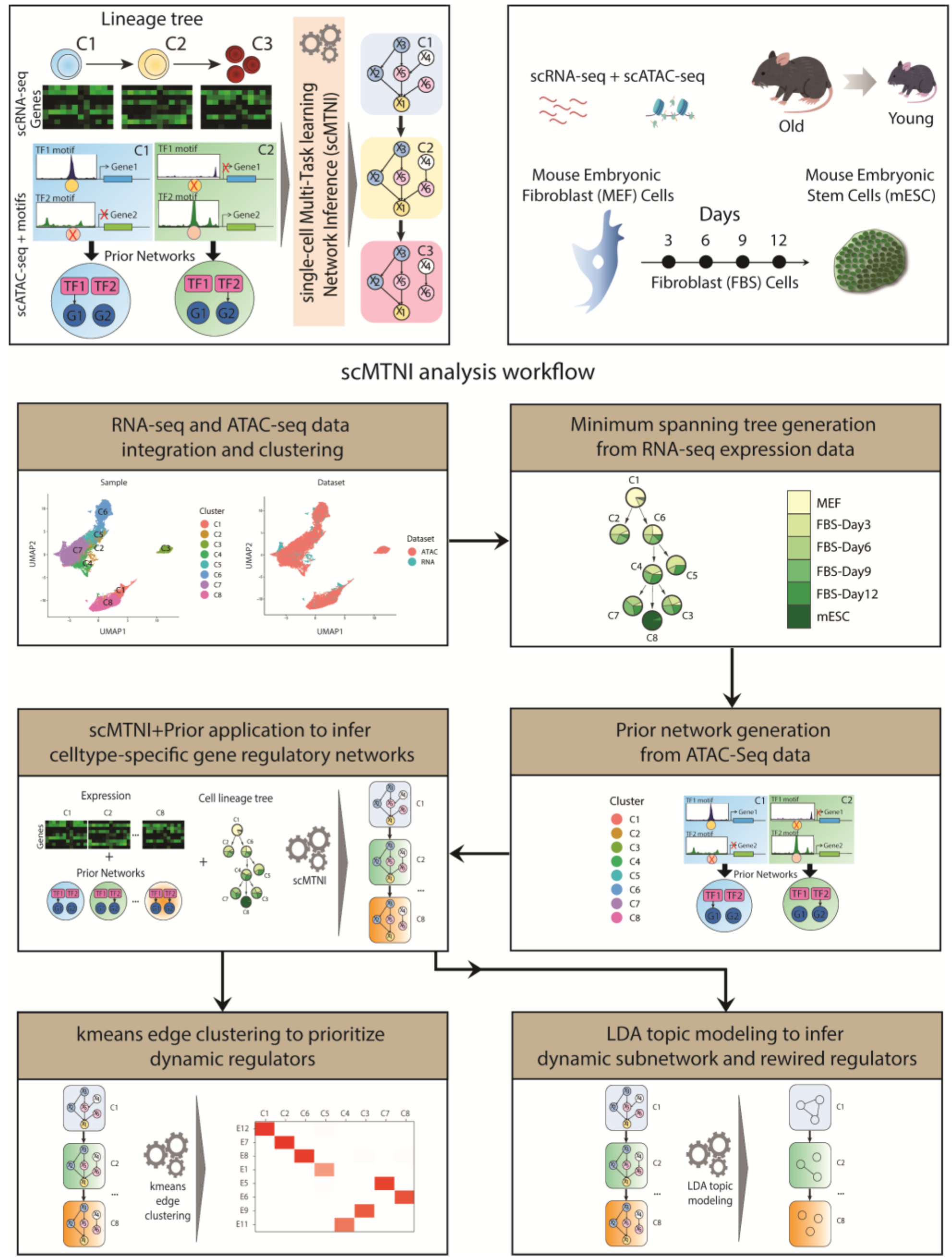

***Figure 1: Overview of the scMTNI framework and application to mouse cellular reprogramming.*** *The single-cell multi-task network inference (scMTNI) framework integrates single-cell RNA sequencing (scRNA-seq) and single-cell ATAC sequencing (scATAC-seq) data to infer cell-type-specific gene regulatory networks (GRNs) along a lineage. Inputs include (a) cluster-specific scRNA-seq expression matrices, (b) a lineage tree of cell state transitions, and (c) optional prior regulatory information from scATAC-seq-derived transcription factor (TF) motif accessibility. scMTNI is demonstrated on a mouse reprogramming dataset covering six stages, from embryonic fibroblasts (MEFs) through fibroblast serum (FBS)-cultured intermediates (Days 3, 6, 9, and 12) to embryonic stem cells (mESCs). The following are the steps in the scMTNI approach: (i) integrating scRNA-seq and scATAC-seq using joint LIGER clustering; (ii) building a minimum spanning tree using lineage*

*information from RNA expression profiles; (iii) creating a prior network using TF-motifs from ATAC profiles; and (iv) determining cluster-specific GRNs. Post-inference analyses apply (v) k-means edge clustering to prioritize dynamic regulators and (vi) latent Dirichlet allocation (LDA) topic modeling to detect subnetworks and rewired regulators driving pluripotency.*

# 4. Methods

To perform scMTNI, the main four steps are: downloading and installing scMTNI (subsection **4.1**), preparing input files for scMTNI (subsection **4.2**), application of scMTNI to mouse cellular reprogramming dataset (subsection **4.3**), evaluation and validation of results (subsection **4.4**).

## *4.1. Downloading and installing scMTNI*

scMTNI is freely available from the Roy Lab GitHub repository: https://github.com/Roy-lab/scMTNI/tree/master. The current implementation of scMTNI is supported on Unix-based operating systems. The GitHub distribution of scMTNI includes the essential GNU Scientific Library (GSL) required for its execution. The following Unix commands can be used to download and install scMTNI via the command-line terminal.

```
git clone https://github.com/Roy-lab/scMTNI.git
cd scMTNI/Code/
make
```

After compilation is complete, the main scMTNI executable will be available as `scMTNI/Code/scMTNI.`

## *4.2. Preparing inputs for scMTNI*

scMTNI requires a set of input files in a specific format. The steps for creating these files are detailed below. The input files and scripts used to generate them are available at: https://github.com/Roy-lab/scMTNI/tree/master.

### *4.2.1. Configuration file*

This file specifies the paths to the cell cluster-specific input data files and the output directories. Each line corresponds to a cell cluster and should contain the following tab-delimited entries in the following order: (1) cell cluster name, (2) expression file name, (3) output directory, (4) regulator list file, (5) target gene list file, (6) prior network file. Below is an example of the configuration file, for three example cell types, C1, C2 and C6. Here `results` is the top-level output directory. All input files other than the prior networks are in the `data` folder. Prior networks are assumed to be in the `prior` directory. We now describe the files individually.

```
C1 data/C1.table results/C1 data/C1_allregulators.txt
data/C1_allGenes.txt prior/C1_network.txt
```

```
C2 data/C2.table results/C2 data/C2_allregulators.txt
data/C2_allGenes.txt prior/C2_network.txt
C6 data/C6.table results/C6 data/C6_allregulators.txt
data/C6_allGenes.txt prior/C6_network.txt
```

### *4.2.2. Cell cluster-specific expression data file*

The expression data file for each cell cluster contains genes as rows and cells as columns. Preprocessing steps for the expression data are described in **note 1**. For each cell cluster, gene names are appended with a suffix indicating the cluster (e.g., in the file `C1.table`, the gene names appear as `genename_C1`). Below is an example of the expression data for the C1 cell cluster, for 4 cells and three genes.

```
Gene	bc-1	bc-2	bc-3	bc-4
Acaca_C1	0.0	6.87908	0.0	0.0
Agtrap_C1	1.36922	0.0	7.12018	0.0
Amacr_C1	0.0	4.53725	0.0	0.0
Chd1_C1	0.0	0.0	0.0	7.30254
```

### *4.2.3. Cell cluster-specific regulator and target list file*

The cell cluster specific scMTNI input files, such as the regulator list and target gene list, are generated using a utility Python script, `PreparescMTNIinputfiles.py`, available in the GitHub repository.

```
python PreparescMTNIinputfiles.py --filelist $filelist --regfile
$regfile --indir $indir --outdir $outdir --splitgene 50 --motifs 1
```

To run this script, two inputs are needed, (1) `filelist`, (2) `regfile`. Here `filelist` specifies the cell cluster names and the corresponding expression files. For example, for our three cell type example, the first three lines of `filelist` is:

Here `--motifs` argument is set to 1 to indicate that the study incorporates prior networks derived from motif information. If motif information is not used, this argument is set to 0.

```
C1	data/C1_expression.txt
C2	data/C2_expression.txt
C6	data/C6_expression.txt
```

The `--splitgene` argument specifies the number of gene sets that the input should be split into to enable parallelization of the inference task. The above usage specifies this to be 50.

Finally, the `--inputdir` argument specifies the top directory for the input files in filelist.txt and `--outputdir` specifies the location of the output files.

Running the script generates a set of files and subfolders within the data directory, including the regulator and target gene list files, which should be readily usable for scMTNI. For example, a cell cluster-specific regulator file (e.g., `C1_allregulators.txt`) looks like the following:

```
Ahctf1_C1
Ahr_C1
Ahrr_C1
Alx1_C1
```

In addition, the script creates a subfolder named `ogids`, which contains split target gene files, the number of files controlled by the `--splitgene` argument. Each target gene file is named sequentially from `AllGenes0.txt`, `AllGenes1.txt`. The maximum number of such files is controlled by the total number of input genes and splitgenes argument. In our example of 14,953 genes, we have a total of 300 (~14,953 ÷ 50) target gene sets.

Notably, three additional files are generated in this step: `testdata_ogids.txt`, `TFs_OGs.txt` and `AllGenes.txt`. The file testdata_ogids.txt contains ortho group information for genes. Each row represents one ortho group. The first column uses the format `OG{ID}_1`, where ID is a numeric identifier for the gene group. The second column lists gene names corresponding to that ortho group, with entries separated by commas and spanning all cell clusters. An example from `testdata_ogids.txt` for three cell types, C1, C2, C6 is shown below:

```
Gene_OGID    NAME
OG1_1 0610007P14Rik_C1,0610007P14Rik_C2,0610007P14Rik_C6
OG2_1 0610009B22Rik_C1,0610009B22Rik_C2,0610009B22Rik_C6
.
.
OG14952_1    Zzz3_C1,Zzz3_C2,Zzz3_C6
OG14953_1    l7Rn6_C1,l7Rn6_C2,l7Rn6_C6
```

The file `TFs_OGs.txt` contains the numeric group IDs corresponding to transcription factors, while `AllGenes.txt` contains the numeric group IDs corresponding to all the target genes.

#### *4.2.4. Motif prior network*

The motif-based prior networks can be cell-type-specific, if scATAC-seq is available, or can be cell type agnostic leveraging either bulk ATAC-seq or sequence-specific motifs alone. scMTNI can fully exploit cell-type-specific priors and we describe now how this can be generated for unpaired datasets with both scRNA-seq and scATAC-seq measurements available but not for the same cell. If paired data is available, for example, from 10x multi-ome, the scRNA-seq clusters can be used to define scATAC-seq clusters.
The current scMTNI workflow uses ArchR to process scATAC-seq to obtain gene scores and LIGER [16] to integrate scRNA-seq and scATAC-seq gene score matrix data (see also **note 2**). The ATAC genescore matrix follows a similar format as the expression matrix. We provide `LIGER_scRNAseq_scATAC.R` utility script to perform LIGER-based integration, however, other integration tools can be used well (**Figure 2a-d**). Application of this workflow to our reprogramming dataset provided a total of eight cell clusters, designated as C1-C8 (**Figure 2a**). We set the number of cell clusters to 8 based on the analysis in the original scMTNI paper by Zhang *et al.* (2023), which tested different k values (8, 10, 12, 15 and 20) and found k=8 to provide the best integration of RNA and ATAC seq data. We note here that scMTNI was applied on the entire set of eight clusters, though we recommend the user to consider merging or excluding very small clusters, e.g, less than 20 cells, which would exclude C5 in our analysis.
With the clusters in hand, we use ArchR to obtain cell cluster-specific ATAC peak files using the `addReproduciblePeakSet` function. The resulting peak files for each of the eight clusters are saved in BED format and are used to generate the motif-based prior networks (see https://zenodo.org/records/16914547).
Using these peak files, we followed steps 3 to 7 from the scMTNI prior network generation pipeline (can be found at: https://github.com/Roy-lab/scMTNI/tree/master/Scripts/genPriorNetwork) to construct the motif-based prior networks. An example of one of the resulting tab-separated prior network files (`C1_network.txt`) is shown below:

```
Ahr_C1      Cyp1a1_C1   1
Ahrr_C1     Cyp1a1_C1   1
Arid3a_C1   Slc35g1_C1 1
Arid3b_C1   Slc35g1_C1 1
Arid3c_C1   Slc35g1_C1 1
```

#### *4.2.5. Specifying the cell lineage tree and cell type order*

scMTNI assumes the cell lineage tree is provided as input. However, in cases where this information is not available, scMTNI uses the PAGA tool [17] on the scRNA-seq clusters to obtain trajectory information followed by extraction of a minimum spanning tree (MST) from PAGA abstractions.

Next, based on the PAGA MST edge list, we generated the cell lineage tree file required as input for scMTNI. The lineage information is provided as a tab-separated text file with four columns: (1) child cell cluster, (2) parent (predecessor) cell cluster, (3) branch-specific gain rate ($p_g$) (4) branch-specific gain rate ($1 - p_m$). The branch-specific gain rate is the probability that a regulatory edge is gained in a child cell type/cluster if it was absent in the parent. The branch-specific loss rate is the probability that an edge is lost in the child cluster if it was present in the parent (see **scMTNI theory**). For this study, the cell lineage file, referred below as `celltype_tree_ancestor.txt`, was manually created using the PAGA MST edge list and the scMTNI structure prior probabilities ($p_g$=0.2 and $p_m$=0.8). An example of the file is shown below:

```
C2    C1    0.2   0.2
C6    C1    0.2   0.2
C5    C6    0.2   0.2
C4    C6    0.2   0.2
C3    C4    0.2   0.2
C7    C4    0.2   0.2
C8    C4    0.2   0.2
```

With the lineage tree in hand, we obtain the cell type order based on a tree traversal from the root. This is specified in the file `celltype_order.txt` (**Figure 2i**), as shown below:

```
C1
C2
C6
C5
```

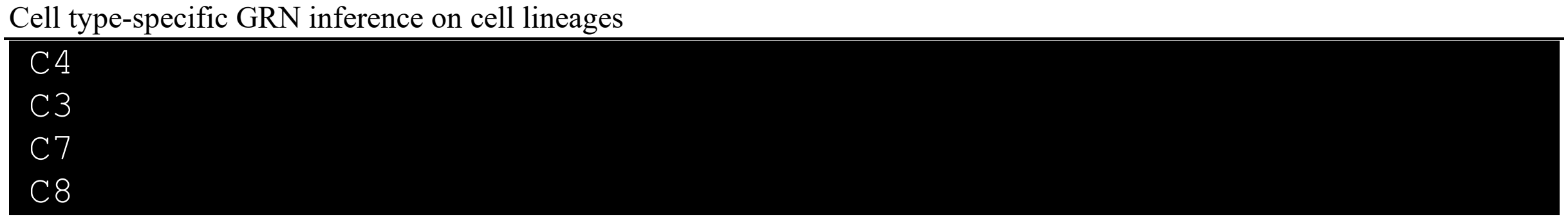

```
C4
C3
C7
C8
```

The `celltype_order.txt` file contains the list of all cell clusters that have data for which GRN inference will be done. The order of clusters in `celltype_order.txt` must match the order of cell clusters listed in the second column of the OGID file.

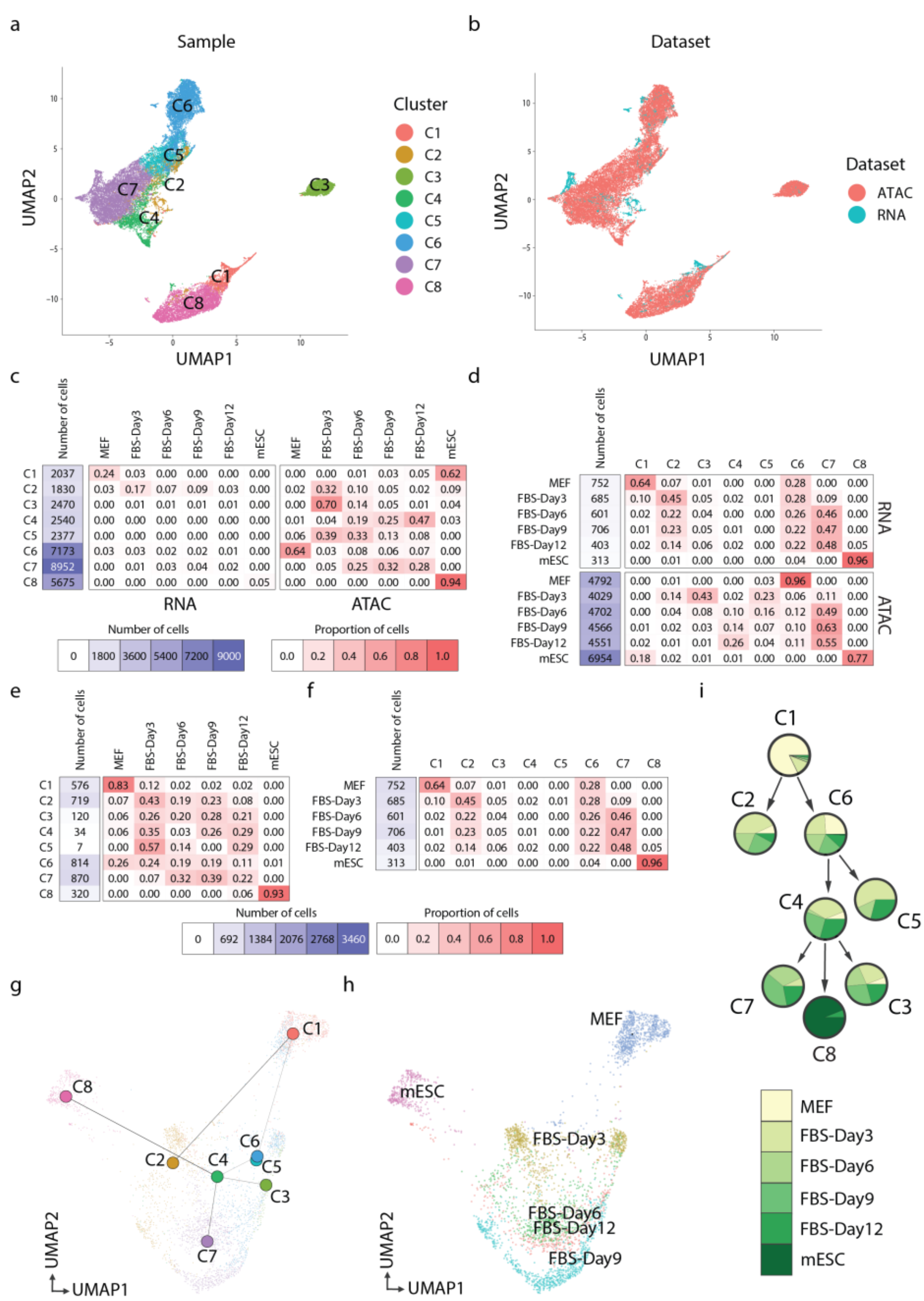

***Figure 2: LIGER-based integration of scRNA-seq and scATAC-seq data, and PAGA-inferred minimum spanning tree constructed from RNA expression profiles. (a)*** *UMAP of LIGER cell*

*clusters on the scATAC-seq data and scRNA-seq data. **(b)** UMAP depicts the sample labels of the scATAC-seq and scRNA-seq data from mouse cellular reprogramming. **(c)** The distribution of both RNA and ATAC samples in each LIGER cell clusters. **(d)** The distribution of LIGER cell clusters in each RNA and ATAC sample. **(e)** The distribution of only RNA cells in each LIGER cell cluster. **(f)** The distribution of LIGER cell clusters only in RNA samples. **(g)** PAGA-inferred minimum spanning tree for scMTNI connecting the eight cell clusters (C1-C8) obtained from RNA expression **(e and f)** data.*

### *4.3. Applying scMTNI*

#### *4.3.1 Usage and outputs of scMTNI*

An example usage of scMTNI run using the example data is as below assuming it is installed in `scMTNI/Code/scMTNI`:

```
scMTNI/Code/scMTNI -f config.txt -x50 -l TFs_OGs.txt -n AllGenes{ogid}.txt -d celltype_tree_ancestor.txt -m testdata_ogids.txt -s celltype_order.txt -p 0.2 -c yes -b -4 -q 1
```

The description of each argument in the scMTNI run is as follows:

- ***-f*** config file with six columns, rows for each cell. Each cell's row should have the following species-specific entries (see **Section 4.2.1**).
- ***-x*** specifies maximum number of regulators to be used for a given target. Here we have used 50.
- ***-l*** list of the group id #s to be considered as regulators (see **Section 4.2.3**).
- ***-n*** List of the group id #s to be considered as potential targets (see **Section 4.2.3**). This option can be used to run scMTNI in parallel (see **note 4**).
- ***-d*** the cell lineage tree to be used (see **Section 4.2.5)**.
- ***-m*** a file that describe the gene relationships (see **Section 4.2.3**).
- ***-s*** cell types present in the gene relationships file (from the parameter *-m*), in the order they exist in the gene file (see **Section 4.2.5**).
- ***-p*** the probability that an edge is present in the root cell. Default value is 0.5 (see **scMTNI theory**).
- ***-b*** Sparsity penalty ($\beta_0$) (see **note 3**).
- ***-q*** Prior penalty ($\beta_1$) (see **note 3**).

scMTNI outputs cell cluster-specific regulatory edges between regulators and target genes, which are available in the result folder. The output folder structure of the present study contains specific subfolders for each cell cluster, labeled C1-C8. Each cell cluster-specific folder includes a subfolder named `fold0` and a file containing the data likelihood of the scMTNI model, `scoreFile.txt`. Inside the fold0 folder, there are two files: the first is the model parameters output file, `modelparams.txt,` and the most important file is `var_mb_pw_k50.txt`, which contains the inferred regulatory network for each cell cluster. The format of the regulatory network file is similar to the input prior network, with the first column specifying the regulator, the second column the target gene, and the third column represents the regression coefficient. Example lines from the inferred regulatory network file `var_mb_pw_k50.txt` for the C1 cell cluster are as follows:

```
Tcf4_C1     1110065P20Rik_C1 -0.201389
Foxm1_C1    1190002F15Rik_C1 0.212517
```

The absolute value of the coefficient corresponds to the regression edge weight, where a larger value indicates a greater influence of the regulator on the target gene. However, as described below, scMTNI is run in stability selection mode which is recommended for all GRN inference algorithms in which case the value of the regression coefficient is ignored and instead the confidence (described below) is used to assess the effect of the regulator on the target gene.

### *4.3.2 Running scMTNI in stability selection mode*

As with GRN inference algorithms, scMTNI is recommended to be executed within a stability selection framework, which involves generating multiple subsamples of the input data.

We provide a utility script `Datasubsample_sc_merged.py` that can be used to generate data subsamples for scMTNI's multi-task learning framework, available at the GitHub repository. This script accepts the list of expression datasets (`filelist`) and the number of subsamples (`nseed`):

```
python Datasubsample_sc_merged.py --filelist $filelist --indir
$indir --nseed 100
```

This produces a set of cell cluster-specific input expression matrices, with each matrix corresponding to a cluster listed in the `filelist` and saved with the `.table` extension. Each subsample has 2/3rd number of cells as per the original article but can be varied by changing `frac` parameter in the script.

After scMTNI has been applied to all subsamples, we estimate the edge confidence as the fraction of times that edge appears in the inferred networks across all 100 subsamples. The consensus network for each cell cluster can be generated using the script `Makeconsensusnetwork.sh,` available in the GitHub repository with the following usage:

```
nseed=100 # number of subsamples
maxReg=50 # number of regulators
cellfile= celltype_order.txt # cell file order file path
indir=results/ # scMTNI results directory path
nogid=300 # number of ogids

bash Makeconsensusnetwork.sh $nseed $maxReg $cellfile $indir $nogid
```

This script expects the scMTNI output to be organized in the following directory structure: `randseed{0..nseed}/ogid{0..nogid}/Results.tar.gz`.

For downstream interpretation of results, each cell cluster-specific consensus GRN can be filtered using either top X (e.g. X=1000) edges sorted based on confidence (cf) or a threshold on the confidence, (e.g., cf>=0.8). In this study, we extracted both the top 1,000 edges (top=1000) and edges with at least 80% confidence (cf>=0.8) for downstream analyses. Networks using either approach can be extracted using the R script `extractTopconsensusedges.R`, available in the GitHub repository as follows:

```
extractTopconsensusedges=extractTopconsensusedges.R
cf1=cf
cf2=top
n1=0.8
n2=1000
```

```
outputpath=results/analysis
mkdir ${outputpath}
Rscript --vanilla $extractTopconsensusedges $cf1 $n1 ${outputpath}
Rscript --vanilla $extractTopconsensusedges $cf2 $n2 ${outputpath}
```

This process outputs two files for each of the eight cell clusters (C1-C8):
(1) consensus_edges_filteredlowexpression_top1000.txt: cell cluster-specific network with the top1000 edges sorted by confidence.
(2) consensus_edges_filteredlowexpression_cf0.8.txt: cell cluster-specific network with edges with confidence of 0.8 or higher.

Both these networks, per cell cluster, can be used in downstream analyses using *k*-means and LDA analysis to uncover cluster-specific network rewiring and to prioritize key regulators along the lineage.

## *4.4.Dynamic network analysis of scMTNI inferred GRNs*

scMTNI's framework provides two ways to examine GRN dynamics: *k*-means clustering and Latent Dirichlet allocation-based topic modeling. We now describe these in detail. These steps assume scMTNI has been executed in stability selection mode.

### *4.4.1. k-means edge clustering to prioritize dynamic regulators for the edge clusters and its matched cell clusters*

We applied *k*-means clustering to the top 1,000 consensus edges by generating an edge × cell cluster confidence matrix, where each entry represents the scMTNI-inferred confidence (cf) score of an edge in each cell cluster. The clustering was performed using the MATLAB script `StablityKmeansClustering.m`, available in the scMTNI GitHub repository, with parameters n=1000 and *k*=1-30. For each *k*, the script runs *k*-means with 100 replicates, computes the silhouette index to evaluate clustering quality, records within-cluster variance to calculate the cumulative percentage of variance explained, and outputs cluster assignments along with heatmaps of edges across cell clusters. To select the number of clusters, we use the silhouette index, where it is maximized favoring a lower *k* while tolerating a small decrease in the silhouette index if needed. In our demonstration, we selected *k* = 25 based on a discontinuity in the silhouette index, which indicated the optimal number of clusters.

To visualize dynamic regulators across edge clusters and their associated cell clusters, we use another utility script a wrapper script `kmeans_bubbleplot_wrapper.sh`, which internally calls the R script `bubbleplot.R`, also available in the GitHub repository, to visualize the top 5 Regulators per cell cluster. Both scripts are available in scMTNI GitHub repository. *k*-means was applied using the following commands:

```
# Apply k-means for k=1-30 to the results with top1000 consensus
edges:
matlab -nodisplay -nosplash -nodesktop -r
"StablityKmeansClustering; quit()"
# Command line arguments
```

```
# 1 [start of kmeans range i.e. 1]
# 2 [end of kmeans range i.e. 30]
# 3 [path to cluster dir]
# 4 [path to output dir]
# 5 [edge threshold i.e. top1000]
# 6 [prefix i.e. kmeans]
bash  Scripts/Network_Analysis/kmeans_bubbleplot_wrapper.sh  1  30
results/analysis/kmeansclustering_top1000/
results/analysis/kmeansclustering_top1000/ top1000 kmeans
```

We selected $k = 25$ based on a discontinuity in the silhouette index, which indicated the optimal number of clusters. For instance, edge cluster E8 (y-axis in **Figure 3a**) showed strong alignment with cell cluster C8 (x-axis), sharing 85% of its edges, resulting in a high matching score of 0.85 (left panel). E8 contains 926 edges (right panel), and its top-ranked regulators include Tfdp1, Pitx2, Esrrb, Zfp42, and Klf2 (**Figure 3b**). These regulators are consistent with known drivers of pluripotency and mESC identity as supported by previous studies [18–22]. Interestingly, our *k*-means edge clustering identified Tbxt (Brachyury) as the overall highest-degree regulator (degree:162) that is present in edge-cluster E3, which corresponds to a small cell cluster C4 (**Figure 3a**) of 34 cells and is composed of early FBS (Day3) cells (35%) with persistence into Day 9–12 (26-29%) but absent in endpoint mESCs (**Figure 2e**). Tbxt is an important regulator of early embryonic development [23] and mesendoderm specification [24], and its emergence here likely reflects a transient mesendoderm-like state [25]. This could reflect a lack of suppression of the muscle program [2] or a diversion into an alternate lineage pathway, both of which can be barriers to successful reprogramming to the pluripotent state [10]. Overall, our *k*-means edge clustering application on the scMTNI-inferred consensus networks identified well-known pluripotency regulators and revealed novel context-specific regulators that may act as critical modulators of reprogramming trajectories.

#### *4.4.2 Topic modeling to infer dynamic subnetworks and rewired regulators across cell clusters*

We next applied Latent Dirichlet Allocation (LDA) topic modeling on the consensus edges with a confidence score > 0.8 to visualize dynamic subnetworks of both static and rewired regulators across cell clusters. This approach provides a complementary perspective to the *k*-means approach by leveraging LDA's ability to uncover topics, corresponding to latent regulatory patterns.

LDA models are traditionally used for document clustering, where each document is represented as a distribution over topics ($\theta$), and each topic is a distribution over words ($\beta$). In the context of gene regulatory networks, we adapted this framework such that regulators are treated as documents and target genes as words. This formulation enables the inference of both static and dynamic regulators across the developmental cell clusters, where each topic can represent a putative biological pathway [26].

We followed the standard LDA pipeline available in the scMTNI GitHub repository, using the following commands:

```
# Step-1: Apply LDA to subsampled results
matlab  -nodisplay  -nosplash  -nodesktop  -r  "addpath('scripts');
LDA_analysis('results                    /analysis/lda_TFcellbygene/',
'celltype_order.txt',   10,   '_filteredlowexpression',   '_cf0.8',
'testdata', 0); quit()"
# Step-2: Obtain the giant component of the inferred networks
Rscript --vanilla plotNetworks_LDA_cf0.8.R 10
```

```
# Step-3: Extract giant components
bash getGiantComponents.sh
# Step-4: Identify top regulators per component
bash getTopRegPerComponent.sh
# Step-5: Generate network figures per topic
Rscript makeAllGraphs.R
# Step-6: Generate regulator bubble plots per topic
Rscript makeTopicRegBubble_ggplot.R
```

For this study, we used 10 topics to explore regulatory dynamics across the mouse cellular reprogramming continuum, from MEFs to mESCs (**Supplementary Figures 1**, **2**). Each topic depicted subnetworks that were either unchanged across cell clusters or exhibited gain/loss of edges across cell clusters. For example, among all the topic-cluster subnetworks (10 topics x 8 cell clusters, C1-C8) (**Supplementary Figure 1**), Ybx1 uniquely emerged from Topic 9 as a rewired regulator, gaining and losing targets as evidenced by the bubble plot size, and was consistently represented across all cell clusters (**Figure 3c-d**, **Supplementary Figure 2**). Ybx1 has been shown to have a dynamic regulatory role during cellular reprogramming, where it modulates RNA stability and translation to fine-tune gene expression during state transition [27]. In line with expression patterns, MEF-associated regulators such as Runx2 (Topic 4) showed prominent activity in early clusters, highlighting the persistence of fibroblast identity. In contrast, no hub regulator expression was uniquely restricted to the ESC state. A total of 49 regulators (**Supplementary Figure 2c**) from 10 different topic networks (**Supplementary Figure 1**) uncovered here is reported to orchestrate essential biological functions underlying mouse cellular reprogramming, ranging cell cycle progression (E2f7/8, Foxm1 from Topic 10) [28], stress adaptation (Atf4, Trp53 from Topic 7) [29], lineage specification for embryonic development (Gata6 and Pax3 from Topic 5, Hoxc6 from Topic 3) [30] and mesenchymal fate and function (Prxx1/2 from Topic 6 and 3) [31]. Interestingly, Klf4 appeared as a top regulator from Topic 2 network that targets nine genes (e.g., Esrrb, Sox2, **Supplementary data**), important for maintenance of the pluripotency regulatory network [32], and found to be primarily associated with C8 cells that mostly represent mESC cells (93%, **Supplementary Figure 2b**). Klf4 (Krüeppel-like factor 4) is one of the Yamanaka reprogramming factors, along with Oct4, Sox2, and c-Myc, that promote pluripotency [33] and drive cellular reprogramming through chromatin remodeling [34] and mesenchymal-to-epithelial transition [35].

This multi-step pipeline allowed us to identify topic-specific subnetworks, visualize key regulators per topic, and characterize regulatory rewiring events across the cell clusters. Each topic provides unique insights into the regulatory dynamics underpinning cellular transitions during reprogramming.

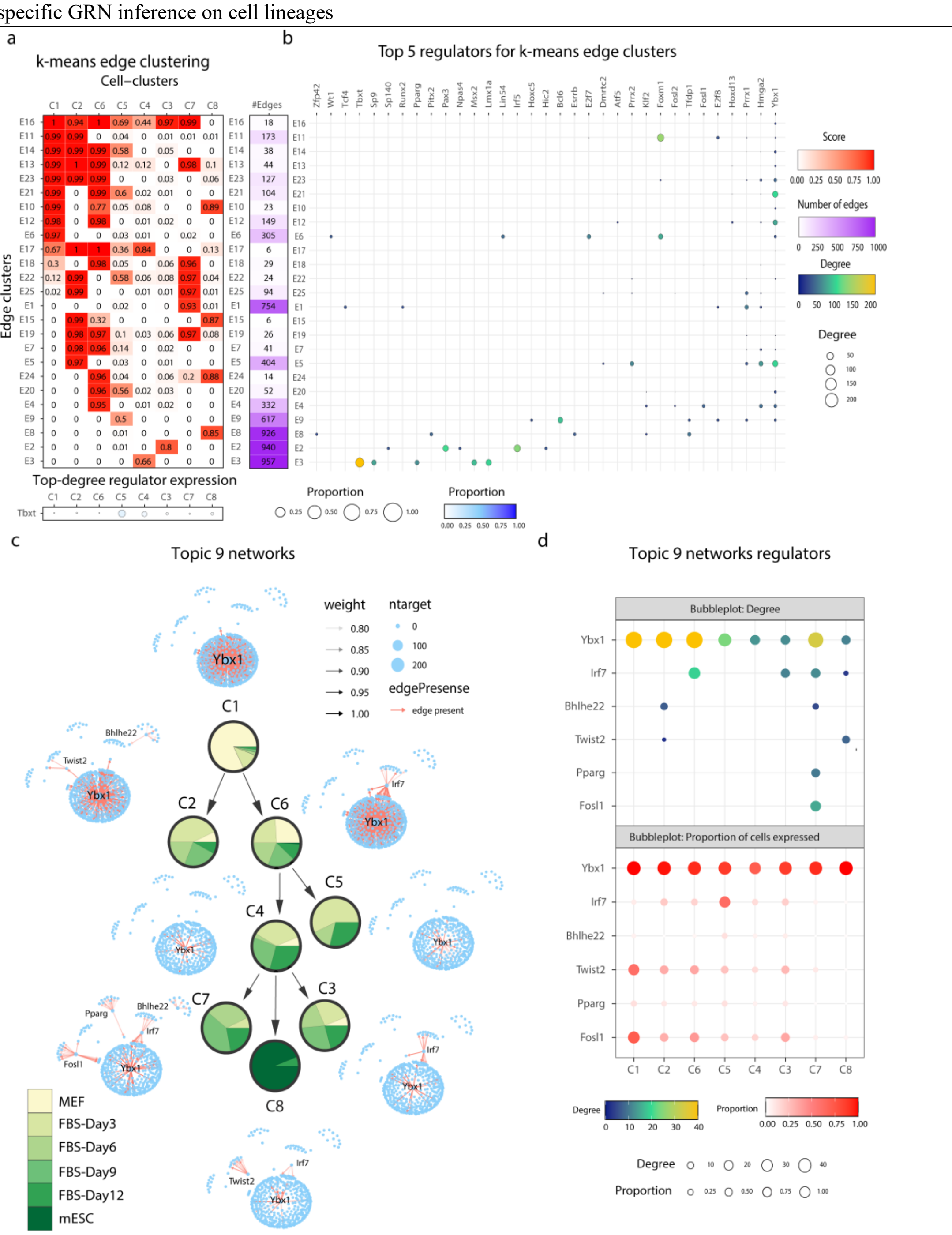

*Figure 3:* ***Dynamic network analysis of scMTNI-inferred GRNs. (a)*** *k-means edge clustering of the top 1,000 high-confidence edges in scMTNI networks. Left: heatmap of edge confidence across 25 edge clusters (E1-E25, rows) and eight cell clusters (C1-C8, columns), colored white-red (low-high, range 0-1). Right: heatmap of edge counts per cluster, colored white-violet (0-1,000).* ***(b)*** *Bubble plot of the top five regulators per edge cluster (≥10 targets). Bubble size and color denote regulator degree, scaled continuously (blue-green-yellow). Per-cell-cluster expression of the top regulator Tbxt (Brachyury) is shown as a bubble plot, with bubble color (white to blue) representing the expression level and bubble size indicating the proportion of cells expressing the gene (0-1).* ***(c)*** *Latent Dirichlet*

*allocation (LDA) topic modeling of edges with ≥0.8 confidence. Topic 9 subnetwork is shown, with red edges and regulators of degree >10.* ***(d)*** *Bubble plots of Topic 9 regulators. Top: node degree (size 0-40; blue-green-yellow gradient). Bottom: proportion of cells expressing each regulator, calculated as number of expressing cells divided by cluster size (size 0-1; white-red gradient). Applied to a mouse reprogramming dataset covering six stages, these analyses highlight dynamic regulators and rewired subnetworks driving pluripotency.*

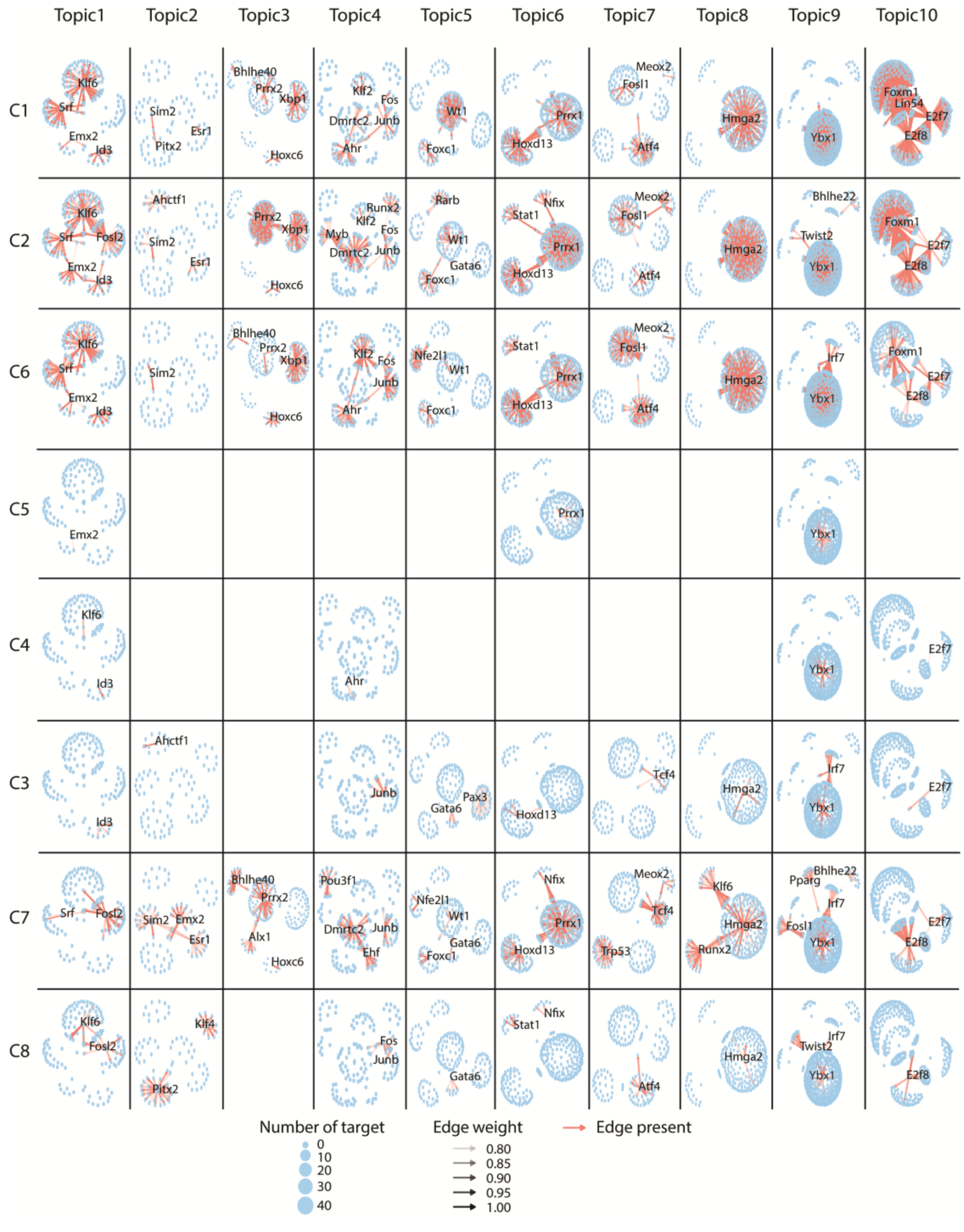

***Supplementary Figure 1. LDA topic-wise per-cell-cluster GRN and inferred rewired regulators identified in cellular reprogramming. (a)*** *Regulatory networks were inferred for each of the 10 topics across the eight cell clusters (C1-C8), resulting in a total of 80 networks. Nodes represent regulators and their targets, with edges (red color) denoting inferred regulatory interactions. Each panel represents the network for a specific topic (presented in columns) in each cell cluster (presented in rows). This illustrates how regulatory connectivity is altered, or 'rewired', across clusters for a given topic, demonstrating the dynamics of the regulator over the course of the cellular reprogramming lineage.*

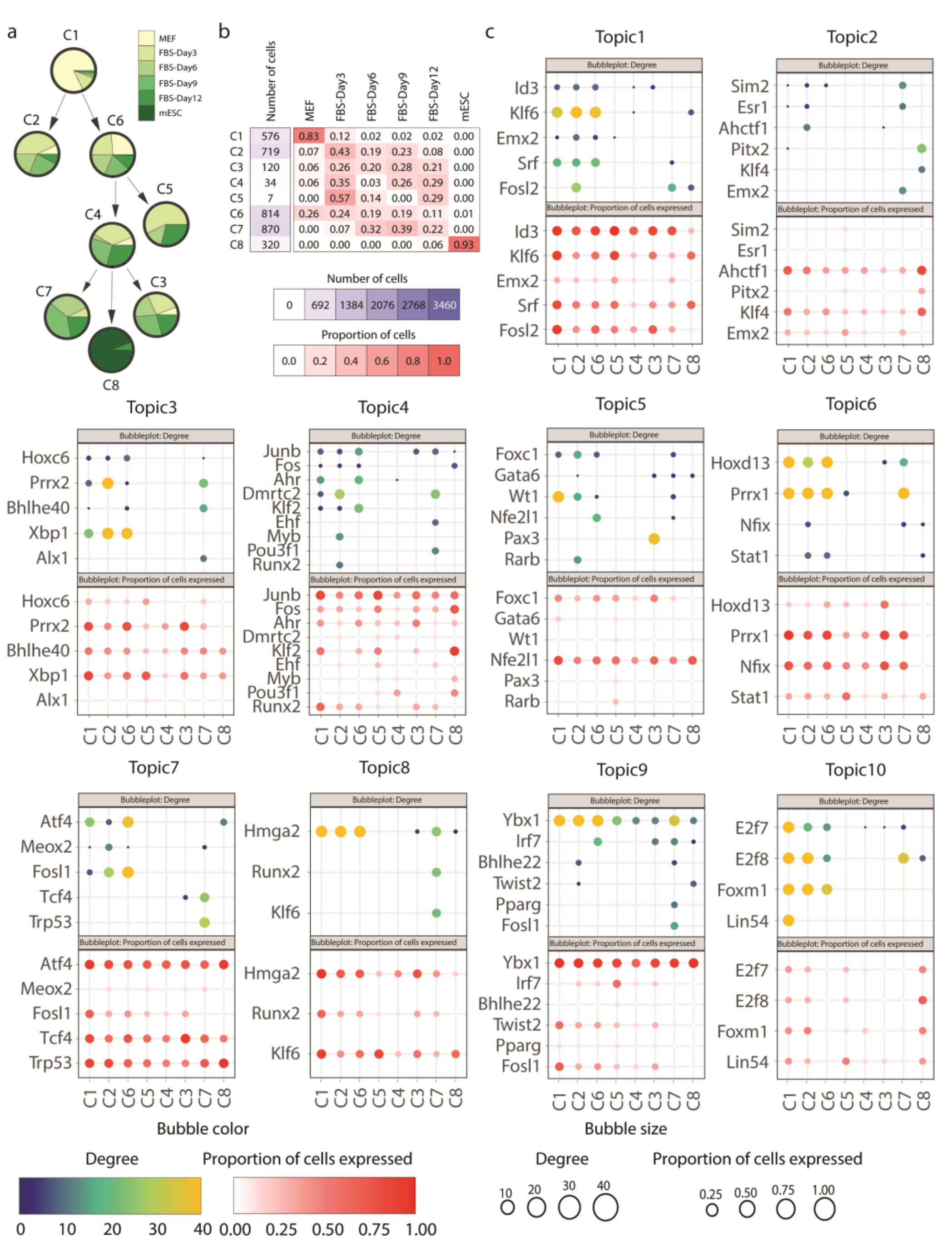

***Supplementary Figure 2: Topic-wise degree distribution and expression bubble plots of LDA-inferred rewired regulators in the mouse cellular reprogramming lineage. (a)*** *Trajectory of mouse cellular reprogramming showing LIGER-defined cell clusters.* ***(b)*** *Distribution of RNA cells across reprogramming cell types where the regulators are expressed.* ***(c)*** *Per-topic bubble plots of rewired regulators across clusters. The top panel shows node degree (size 0-40; blue-green-yellow gradient), and the bottom panel shows the proportion of cells expressing each regulator (number of expressing cells divided by cluster size, size 0-1; white–red gradient) for all 10 topics. Notably, Ybx1 from Topic 9 is the only regulator consistently present across all reprogramming clusters (Figure 3c–d). Other key regulators are discussed in the main text.*

## 5. Notes

### 1. Pre-processing of gene expression data

The input expression data for scMTNI consists of scRNA-seq and/or scATAC-seq data. The scRNA-seq data (GEO ID: GSE108222) in mouse cellular reprogramming data is downloaded for six time points, corresponding to the starting MEF, the end pluripotent state (mESC), and four intermediate timepoints of day 3, day 6, day 9 and day 12 (**Figure 1**). The expression data from two replicates at each time point were merged and normalized for sequencing depth and variance stabilization using pagoda pipeline [36]. For each time point, genes expressed in fewer than five cells were removed. We then took the union of genes across all time points and combined the expression data into a final scRNA-seq matrix. The processed dataset consists of 14,953 genes and 3,460 cells.

To normalize the expression data, we can do log-transformation so that highly expressed genes do not overshadow those with lower expression. In this study, the subsample expression data is log-transformed using the natural logarithm (base e) to reduce variance, address low counts, and manage data overdispersion.

### 2. Generating gene score matrix from ATAC signals

ATAC-seq fragment files were downloaded from NCBI-GEO using the accession number GSE208620 for the same time point as we have scRNA sequencing data is available (**Figure 1**). The fragment files were then processed using the ArchR tool [37] with the latest mouse genome, mm10, to create arrow files using the following thresholds: (i) minimum transcription start site (TSS) enrichment score (minTSS) = 4, (ii) minimum number of mapped ATAC-seq fragments per cell (minFrags) = 1000, and (iii) size of the tiles used for binning ATAC counts (tileSize) = 1000. During the creation of arrow files, two important parameters, addTileMat and addGeneScoreMat, were set to 'TRUE' to extract the gene score matrix. Synthetic doublet cells were removed from the arrow files using UMAP dimensionality reduction, implementing the k-nearest neighbor algorithm at $k$ = 10. After filtering, the gene score matrix had 24,333 genes and 29,594 cells (see subsection **4.2.4** for implementation).

### 3. Setting the parameters: sparsity penalty and prior penalty

Sparsity penalty parameter ($\beta_0$), specified by the `-b` argument, controls the penalty of adding a new edge to the network, and it takes a negative value. We recommend considering a range of $-0.1 \leq \beta_0 \leq$ $-5$. A more negative value increases the cost of adding edges, leading to a sparser network. In this study, we set $\beta_0 = -4$ (see subsection **4.3.1**).

Prior penalty parameter ($\beta_1$), specified by the argument `-q`, controls how motifs are incorporated as priors, so it takes a positive value. We suggest considering a range of $0 \leq \beta_1 \leq 5$. A higher value increases the weight of the motif prior network. In this study, we set $\beta_1 = 1$ (see subsection **4.3.1**).

**4. Running scMTNI in parallel for multiple gene subsets**

To make scMTNI run faster, we can split the target gene set into subsets (e.g., 50 genes per subset) and run scMTNI in parallel on each subset. This can be achieved by dividing the potential target ortho groups into non-overlapping sets of 50 ortho groups per set and storing the corresponding OGID numbers into files such as `AllGenes0.txt`, `AllGenes1.txt`, etc. (see subsection **4.2.3**). Next, scMTNI can be run in parallel using each of the potential target ortho group ID `(OGID)` files (`AllGenes${i}.txt`), while keeping the rest of the arguments the same. After all the parallel runs are complete, we merge the inferred networks for each set of 50 target genes to obtain the final inferred network.

## 5.5 Generating a cell lineage tree from data

In cases where the cell lineage structure is not available, we can use PAGA or a similar lineage approach. We describe below the steps of building a cell lineage tree for the C1-C8 cell clusters.

For this mouse cellular reprogramming dataset, we generated a cell lineage tree connecting clusters C1 to C8 using the minimum spanning tree (MST) methodology, based on the partition-based Graph Abstraction (PAGA) framework applied to the RNA component of integrated RNA+ATAC data (**Figure 2e**, **f**). PAGA constructs a minimum spanning tree that connects all clusters while minimizing the total edge weight as distance, or "1-similarity", by calculating the pairwise similarities using k-nearest neighbor (kNN) algorithm between clusters based on the RNA expression of the clusters' constituent cells [17]. Here, we set kNN=15 to balance sparsity and reproducibility, allowing PAGA to generate an MST-based lineage structure that reflects potential cellular trajectories. We used the Python script `run_paga.py`, which takes an expression matrix, cell cluster information, and a kNN value as input, and produces a matrix representing the cell lineage tree. This script is available on GitHub and can be executed in via the wrapper bash script `apply_PAGA.sh` as follows:

```
run_PAGA=run_paga.py # Declaration of the location of python
executable for PAGA tool
matrix=count_data_dense.txt #input matrix (cells x genes) :: Cells
on rows, Genes on column
cell_file=ligerclusters_RNA.txt #Input cell cluster info file
out_dir=results #output result directory
for i in 15
do
    mkdir -p ${out_dir}/kNN_${i}
    cd ${out_dir}/kNN_${i}
    echo ${i}
```

```
    python3     $run_PAGA     ${matrix}     ${i}     ${cell_file}
${out_dir}/kNN_${i}
    echo
    cd -
done
```

The resulting PAGA MST edge list (**Figure 2g–h**) shows cluster connections, with distances (1 - similarity) in the third column:

```
C1   C2   0.6644218470594079
C3   C4   0.4247111344537815
C1   C6   0.2904516478735228
C4   C6   0.2590356781533252
C5   C6   0.0894173394173394
C4   C7   0.5948964068385976
C4   C8   0.8637473739495798
```

## Acknowledgements

This work is supported by the National Institutes of Health (Grant ID GM144708-01A1), and the Department of Energy (DOE grants DE-SC0021052, DE-SC0023082).

**Supplementary data**

Supplementary data with datasets, results and analyses have been deposited in Zenodo and are accessible at: https://zenodo.org/records/16914547.